\documentclass[sigplan, authorversion]{acmart}

\usepackage{amsmath, amsfonts}
\usepackage{xspace}

\usepackage{orcidlink}

\usepackage{graphicx}
\usepackage{subcaption}

\usepackage[inline]{enumitem}
\setlist*[itemize]{labelindent=10pt, itemindent=0pt, leftmargin=*}

\usepackage{booktabs}
\usepackage{multirow}
\usepackage{tabularx}

\usepackage{wasysym}

\usepackage{pgfplots}
\pgfplotsset{compat=1.18}
\usepgfplotslibrary{statistics}

\usepackage{listings}
\usepackage{iftex}

\newcommand{\listsize}{\fontsize{7}{8}\selectfont}

\ifPDFTeX
  \usepackage[T1]{fontenc}
  \usepackage[utf8]{inputenc}
  \usepackage{zi4} 
\else
  \usepackage{fontspec}
\fi

\definecolor{vscode-bg}{HTML}{FFFFFF}
\definecolor{vscode-fg}{HTML}{1F2328}

\definecolor{vscode-keyword}{HTML}{0000FF}
\definecolor{vscode-string}{HTML}{A31515}
\definecolor{vscode-comment}{HTML}{008000}
\definecolor{vscode-number}{HTML}{098658}
\definecolor{vscode-func}{HTML}{795E26}
\definecolor{vscode-type}{HTML}{267F99}

\definecolor{LightGrey}{HTML}{BFBFBF}

\lstdefinelanguage{CPSLint}{%
      morekeywords={import, from, export, to, csv, is, skip, impute, in, real, nat, int, uart, str, bool, critical, sorted, unique, last, next, mean, median, interpolation, linear, time, zero, quadratic, cubic, barycentric, piecewise_polynomial, pchip, akima, cubicspline, from_derivatives, regex, empty, out, of, order, cut, when, inspect, using, perform},
      morestring=[b]',
}

\lstdefinelanguage{Rascal}{%
      morekeywords={syntax, data, module, alias, list, tuple, lrel, str, real, bool},
      morestring=[b]",
}

\lstdefinelanguage{PythonExtended}[]{Python}{
  morekeywords={as},
}

\usepackage[nameinlink]{cleveref}

\usepackage{csquotes}
\usepackage{textcomp}

\usepackage{balance}

\newcommand{\CPSLint}{\textsf{CPSLint}\xspace}

\copyrightyear{2026}
\setcopyright{cc}
\setcctype{by}

\acmDOI{10.1145/3806383.3815519}
\acmISBN{979-8-4007-2639-2/2026/07}

\acmYear{2026}
\acmConference[SLE '26]{19th ACM SIGPLAN International Conference on Software Language Engineering}{July 02--03, 2026}{Rennes, France}
\acmBooktitle{19th ACM SIGPLAN International Conference on Software Language Engineering (SLE '26), July 02--03, 2026, Rennes, France}

\begin{document}

\title[CPSLint: A DSL Providing Data Validation for Industrial CPS]{CPSLint: A Domain-Specific Language Providing Data Validation and Sanitisation for Industrial Cyber-Physical Systems}

\author{Uraz {Odyurt}}
\orcid{0000-0003-1094-0234}
\affiliation{%
	\institution{Faculty of Engineering Technology, University of Twente}
	\city{Enschede}
	\country{The Netherlands}}
\email{u.odyurt@utwente.nl}

\author{Ömer {Sayilir}}
\orcid{0009-0009-8860-2316}
\affiliation{%
	\institution{Formal Methods \& Tools, University of Twente}
	\city{Enschede}
	\country{The Netherlands}}
\email{o.f.sayilir@utwente.nl}

\author{Mariëlle {Stoelinga}}
\orcid{0000-0001-6793-8165}
\affiliation{%
	\institution{Formal Methods \& Tools, University of Twente}
	\city{Enschede}
	\country{The Netherlands}}
\email{m.i.a.stoelinga@utwente.nl}

\author{Vadim {Zaytsev}}
\orcid{0000-0001-7764-4224}
\affiliation{%
	\institution{Formal Methods \& Tools, University of Twente}
	\city{Enschede}
	\country{The Netherlands}}
\email{vadim@grammarware.net}

\renewcommand{\shortauthors}{U. Odyurt et al.}

\begin{abstract}
Industrial cyber-physical systems generate vast amounts of semi-structured time-series data that require careful preprocessing before they can be effectively used for machine learning applications such as fault detection and identification. Raw sensor datasets are often corrupted or incomplete, making it challenging to develop reliable solutions without proper data preparation and validation. In this paper, we introduce \CPSLint, a domain-specific language for data validation and sanitisation. We present the design, implementation and evaluation of \CPSLint, demonstrating its ability to automatically detect and correct common data corruption patterns while enabling non-programming domain experts to effectively prepare their data for analysis. 
We report evaluation results on a representative dataset, tracking memory consumption and CPU-time for sanitisation activities.
Our approach offers several advantages over traditional methods, including reduced manual effort, guaranteed consistency and broader applicability across time-series datasets and projects.
\end{abstract}

%

\keywords{Domain-Specific Language, Data validation, Data sanitisation, Industrial CPS, Time-series}

\maketitle


\section{Introduction}
\label{sec:introduction}
Sensor data collected from industrial Cyber-Physical Systems (CPS) is seldom well-formatted or error-free. It typically consists of large amounts of time-series data logging the system's status in regular time intervals. Such data collections are not directly consumable by data-centric solutions and workflows. They require sanity checks and preprocessing steps before becoming suitable for Machine Learning (ML) solutions, and have enough variation to justify the lack of one universally applicable solution. Data preparation is needed to both ensure robust and reliable operation of solutions, and to enable functionalities dependent on context-specific preprocessing. As such, prior to the consumption of machine data by different analytic workflows, validation, sanitisation, remediation (where applicable), and structuring is necessary. This is particularly apparent and problematic for anomaly detection and identification, which is a cornerstone of Fault Detection and Isolation (FDI) in industrial contexts, as corroborated in multiple surveys~\cite{Arts:2025:FYMO, Kim:2021:COSP, Tamssaouet:2023:SLFP}.

A common way to execute data preparation is to cover it as part of the data-centric workflow. In other words, we program validation, structuring and so forth, into the algorithms as preceding steps to the workflow itself. This is common practice in the CPS domain. One example is applying sanity checks such as validation of input file format consistency. While functionally effective, the approach can be inefficient. Even considering proper software development practices, much ad hoc and per use-case programming will be required. At the same time, this approach will require the permanent involvement of software engineers alongside domain experts. Expanding on the domain knowledge perspective, the fact that data preparation is aimed at enabling context-specific preprocessing, and that data corruptions often follow historical patterns in relation with the type of system, the role of domain experts will be central. \emph{Not every domain expert is a professional software engineer}.

When it comes to the preprocessing of data, a language closer to domain knowledge can be beneficial, as it allows domain experts to be more directly and decisively involved in the design and implementation processes. We improve the robustness of FDI and similar ML-assisted workflows, by providing a dedicated tool to assist with the task of input data preparation/sanitisation. As a separate step, our tool can also perform data compartmentalisation, resulting in segments referred to as \emph{execution phases}, for workflows that rely on it. Examples of incorrect data are numerous; however, we can already mention instances of corruption such as textual characters in numeric fields, unexpected line separators, missing data fields, or even missing rows. We propose and showcase \CPSLint, a specialised Domain-Specific Language (DSL) built for this task, accessible directly to domain experts.

\CPSLint's main features include type checking and enforcing constraints through validation and remediation for data columns, such as imputing missing data from surrounding rows. More advanced features cover inference of CPS-specific data structures, both column-wise (type and unit) and row-wise (compartmentalisation into execution phases). We demonstrate \CPSLint's features through a proof of concept implementation and provide a comprehensive comparison amongst different approaches for data preprocessing.

\begin{figure}[htbp]
	\centering
	\includegraphics[width=0.8\linewidth]{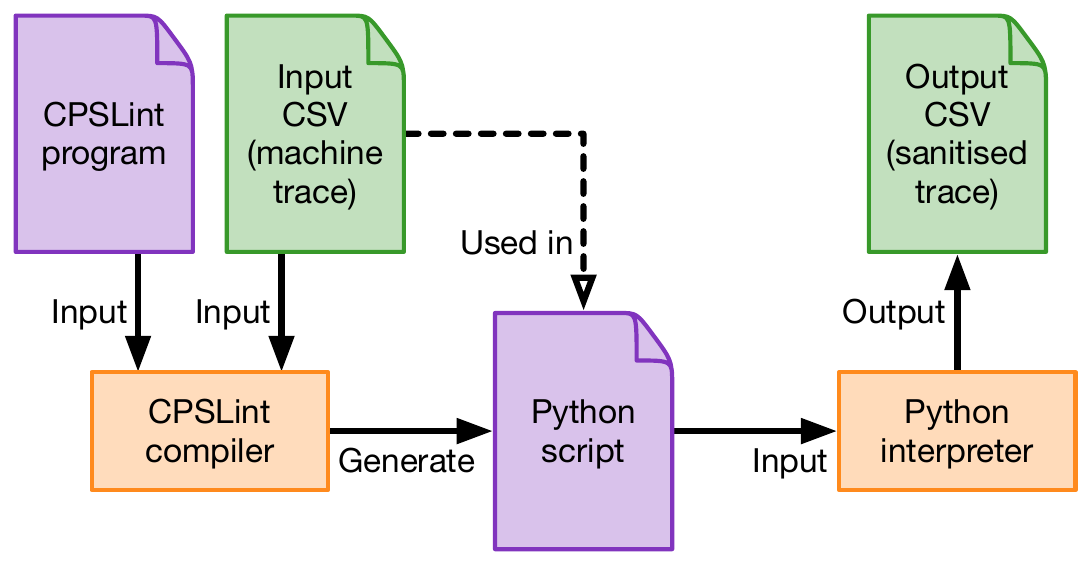}
	\caption{A typical \CPSLint workflow for sanitising CPS machine traces.}
	\label{fig:pipeline_simple}
\end{figure}

\Cref{fig:pipeline_simple} depicts a typical \CPSLint workflow. The compiler takes a \CPSLint specification and a raw CSV file as input, and generates a human-readable Python script that outputs a CSV that is sanitised according to the provided definition.

\begin{figure}[htbp]
	\centering
	\includegraphics[width=0.95\linewidth]{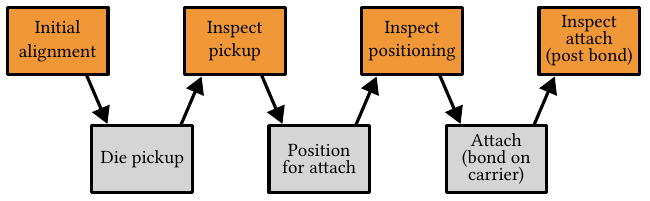}
	\caption{An example CPS machine cycle, involving action steps (in grey) and inspection steps (in orange).}
	\label{fig:machine_cycle}
\end{figure}

\paragraph*{Targeted ML workflows} 
\CPSLint is specifically applicable where time-series data is processed and formed into datasets for ML model input, e.g., FDI. These solutions involve data compartmentalisation using \emph{execution phases} and are primarily intended for processing of machine traces collected during the operation of industrial CPS. Examples include semiconductor photolithography machines, production printing machines, die bonder machines, and similar equipment. A few of the common characteristics apparent in such systems are: presence of highly complex, multi-node compute and control elements; a narrowly defined domain of operational tasks, i.e., highly purpose-built; highly repetitive and predominantly sequential machine cycles; constrained yet non-deterministic behaviour; and a continuous focus on achieving high-yield production output. A die bonder machine from our industrial partner ITEC\footnote{\url{https://www.itecequipment.com/}} is an example of such a repetitive high-yield industrial CPS, with sequential machine cycles per die. A high-level diagram of such machine cycles is drawn in \Cref{fig:machine_cycle}.

\paragraph*{Contributions}
\begin{itemize}
    \item We provide \CPSLint, a software language and a reusable tool dedicated to industrial CPS use-cases, allowing domain experts to perform time-series data preparation, e.g., by cleaning up corrupted fields, or compartmentalising the data. \CPSLint's code is publicly available~\cite{Sayilir:2025:CODE}\footnote{Also available at: \url{https://github.com/omersayilir75/CPSLint}}, and the implementation details are elaborated in a separate document~\cite{Odyurt:2026:ImplementingCPSLint}.
    \item We provide a comprehensive comparison between three approaches for the data preprocessing activity: A \CPSLint approach resulting in compiled Python scripts; another \CPSLint approach with directly interpreted data processing; and the current common practice using ad hoc and Python scripts.
\end{itemize}

After this introduction, background concepts are given in \Cref{sec:background}, followed by the \CPSLint language design details in \Cref{sec:dsl-design}. Our demonstrator is described in \Cref{sec:demonstrator_description} as the evaluation setup with data collection and seeded corruption types. Reflections on the actual usage and different features are given in \Cref{sec:application_of_cpslint}, presenting how the code in \CPSLint looks like. \Cref{sec:benchmarking_setup,sec:discussion} define our benchmarking strategy and discuss the results of its application, respectively. After pointing out the related work in \Cref{sec:related_work}, we share our concluding remarks in \Cref{sec:conclusion}.

\section{Background}
\label{sec:background}

\subsection{Domain-Specific Languages}
DSLs are programming languages with a higher level of abstraction than General-purpose Programming Languages (GPLs)~\cite{Mernik:2005:WHDD}. DSLs are often employed to perform specific tasks, for instance database management languages like SQL~\cite{Chamberlin:1974:SEQUEL}, and to enable domain experts to undertake programming tasks using terms and concepts familiar to them, e.g., MATLAB~\cite{Moler:2020:MATLAB} for scientific computing). The output of DSLs does not always need to be executable; for example, the DOT language of GraphViz~\cite{Gansner:1993:TDDG} can be used to draw graphs, YAML~\cite{Ben-Kiki:2021:YAML} can be used to orchestrate Docker containers using Docker Compose, and \LaTeX{}~\cite{Lamport:1994:LATEX} can be used to write documents. Regardless of the domain or output, DSLs provide a way to perform a task with significantly less effort and expertise than would be required using a GPL. In this paper, we create a DSL leveraging similarities found across the domain of data validation and sanitisation, combined with the information from the cyber-physical domain.

\subsection{Industrial Cyber-Physical Systems}
\label{subsec:industrial_cyber_physical_systems}
Industrial CPS represent a complex integration of physical processes with computational and networked elements, where traditional control methods meet modern data-driven approaches. These systems are characterised by their ability to monitor and respond to changing conditions in real-time, making them fundamental components in industries ranging from manufacturing to healthcare.

This industrial CPS domain encompasses various types of equipment, including but not limited to semiconductor fabrication machines, production printing presses, die bonders, and other high-volume manufacturing tools. What sets these systems apart is their complex nature, combining elements such as multi-node compute and control architectures, highly specialised operational tasks, repetitive yet sequential machine cycles, constrained yet non-deterministic behaviour, and a continuous focus on achieving high-yield production.

\subsection{Execution phases}
\begin{figure}[thbp]
	\centering
	\includegraphics[width=\linewidth]{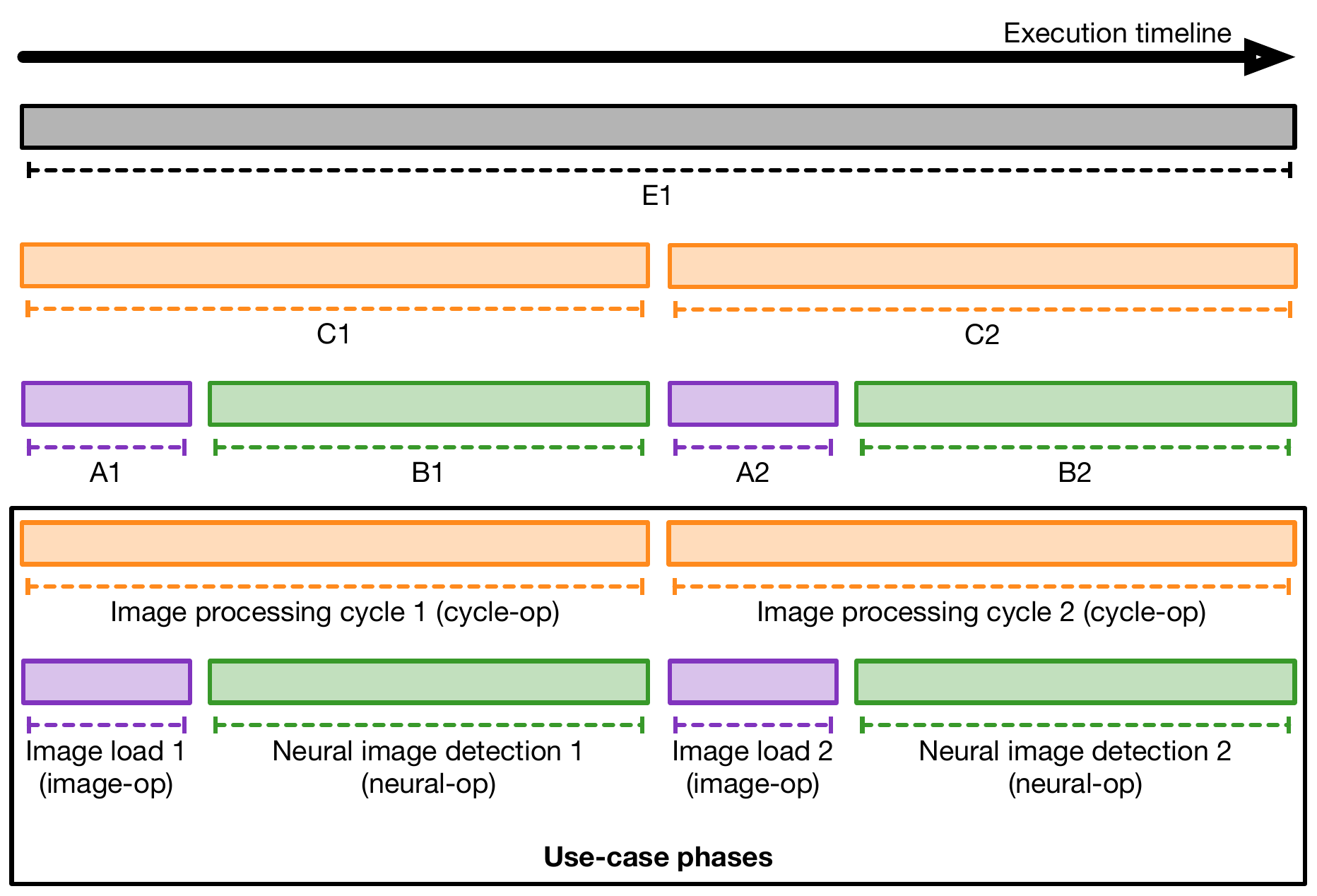}
	\caption{Different data compartmentalisation granularities within an execution timeline, visualising repeated phase types during consecutive rounds of tasks/sub-tasks, plus the phases considered for processing consecutive image data.}
	\label{fig:execution_phases}
\end{figure}

Execution phases are segments of monitoring data, reflecting the behaviour of the system under scrutiny per individual task during an execution. Phases are applicable to time-series monitoring data collected from industrial CPS. As industrial CPS operate primarily on sequential and repetitive machine cycles, the concept of phases is an effective method of data compartmentalisation. As such, data related to each task or sub-task can be analysed in isolation. Phases covering tasks and respective sub-tasks form a hierarchical relation, which can be seen in \Cref{fig:execution_phases} for a generic execution timeline. The bottom two rows depict such a compartmentalisation using phases specific to our demonstrator use-case, described in \Cref{subsec:demonstration_platform}. Considering the repetitive nature of industrial CPS, compartmentalised phase data are most suitable for collecting ML dataset instances.

\subsection{Data preprocessing}
\label{subsec:data_preprocessing}
In cases where ML is used with time-series data from industrial CPS, the concept of execution phases is particularly effective. Preprocessing compartmentalised data based on execution phases results in descriptive features for traditional ML training~\cite{Odyurt:2022:IRIC}. \Cref{fig:execution_phases} visualises the considered phases relevant for our demonstrator, namely the image processing cycle task, made up of the image load and the neural image detection subtasks. As such, the preprocessing should cover sanitisation of machine traces as a preparatory step, followed by trace-cutting, leading to isolated execution phases.

The correct operation of ML-assisted workflows for industrial CPS, or any type of data-reliant system, relies on the correctness of the data. An ad hoc approach is to implement as many sanity checks and corrective actions as possible in the expected presence of corrupt or malformed data. In this context, developing effective preprocessing and performing data validation and sanitisation would benefit from understanding this domain's unique characteristics. With \CPSLint, we aim to bring greater robustness by driving the data preprocessing step of the sanitisation process by providing the domain expert with a DSL that performs this task.

\subsection{Data corruption types}
\label{subsec:data_corruption_types}
Data corruptions are a common occurrence when collecting monitoring data from industrial machines. Whether caused by sensor inaccuracies, data recording subsystem malfunctions, or other culprits, monitoring data collections are accompanied with corruptions. Well-known categories are:

\subsubsection{Data type mismatch}
Disallowed data types are mixed with the correct type, e.g., numeric fields injected with invalid characters (letters or special symbols).
\textbf{Emulation:} Numeric fields are injected with invalid characters, rendering them non-parsable. In some cases, the Universal Asynchronous Receiver/Transmitter (UART) identifier field is also corrupted, simulating noisy channel logs.

\subsubsection{Data type mismatch with targeted UART}
Similar to the previous corruption, but biased towards affected rows associated with a specified UART identifier. This corruption is much more selective and device specific.
\textbf{Emulation:} The UART field is corrupted with invalid characters. UART data fields are especially important for steering the targeted solution and as such, it is important to have cases with noisy UART messages.

\subsubsection{Out-of-bounds values}
Numeric fields that contain extreme constants, which are far outside expected measurement ranges. Depending on the occurrence, this may affect all numeric columns or a random subset within each block.
\textbf{Emulation:} Numeric fields are replaced with extreme constants, e.g., \enquote{99\,999.999}, at random.

\subsubsection{Out-of-order rows with reliable timestamps}
\phantom{xx}
Rows are recorded out-of-order, but their original timestamps are correct. This corruption reorders messages while maintaining temporal information intact.
\textbf{Emulation:} Rows in a block are randomly permuted, but their original timestamps are preserved. This emulates reordering of messages

\subsubsection{Out-of-order rows with unreliable timestamps}
This corruption happens where both order and timing data are compromised during collection.
\textbf{Emulation:} Rows are shuffled and new, monotonically increasing timestamps are generated for the shuffled block.

\subsubsection{Missing fields}
Representing systematic loss of a sensor channel, this corruption could be the case for any column of data.
\textbf{Emulation:} A randomly chosen numeric column is blanked out across all rows in a block, representing systematic loss of a sensor channel.

\subsubsection{Missing rows}
Involves partial loss of data segments, such as dropped packets or truncated logs, resulting in missing rows.
\textbf{Emulation:} Entire blocks of rows are deleted, optionally biased towards containing a specified UART identifier. This models partial loss of data segments, e.g., where important information over UART has to be present.

\subsubsection{Misplaced end-of-line markers}
Involves missing or corrupted line delimiters, which leads to common parsing errors for raw log files.
\textbf{Emulation:} Rows within a block are concatenated with their successors, emulating missing or corrupted delimiters. This results in malformed records and broken row alignment.

\section{DSL design}
\label{sec:dsl-design}
\subsection{Domain}
\CPSLint revolves around the domain of data collected from industrial CPSs. This data is commonly saved in a tabular time-series format, often in databases or \texttt{.csv} files, where each row represents the state of the system in a given point of time. The columns in these files often contain readings from a sensor, such as temperature readings or power usage. These readings can be represented as plain numerical values (natural numbers, integers, real numbers, and so forth), but it would be ideal if they are represented as the unit used for them such as degrees Celsius ($^\circ\mathrm{C}$), amps (A), volts (V), and joules (J).  

While collecting data from industrial CPS, it is not uncommon to end up with cases where fragments of data are either corrupted or lost. This can happen due to, for example, sensor malfunction or human error while preprocessing the data. To get the most out of the data, this data loss and corruption has to be remedied. One way this can be done is by removing invalid readings from the data and imputing the gaps using a fitting imputation policy. Sometimes a simple imputation policy such as copying the previous or next valid value suffices. Other times using statistical properties such as the mean or median can work, and in some cases something more involved such as polynomial interpolation is the appropriate method.

In some cases the output from the preprocessing can benefit from being divided into several datasets, especially for datasets concerning a large period of time. One reason for doing this is to isolate instances of a process by cutting the data into subsets, each covering an instance of the process.

\subsection{Syntax}
The syntax for \CPSLint was designed in close collaboration between language engineers, bringing expertise from the DSL design space~\cite{Mernik:2005:WHDD,Karsai:2014:DGDS,Smaragdakis:2019:NPPL,Zaytsev:2017:LDWI}, and domain experts, contributing knowledge of typical data types, corruption kinds, supported formats, etc. We developed a high-level declarative language with an implementation prototype capable of handling representative CPS data. As shown in \Cref{lst:rascalsyntax}, the language has been kept syntactically simple, revolving around three main actions:
\begin{lstlisting}[
    caption={Concrete syntax for three main actions in \CPSLint.}, 
    label={lst:rascalsyntax}, 
    language=Rascal
]
syntax Action
    = "inspect" FormatName "from" FileName ";"
    | "import"  FormatName "from" FileName ("," {Filter ","}+)? ";"
    | "export"  FormatName "(" {Column ";"}+ ")" "to" FileName "cut" "when" Border "is" Value ";";
\end{lstlisting}

\subsection{Semantics}
Currently the \CPSLint implementation supports a core set of features, providing the following capabilities:
\begin{itemize}
    \item Inspect a dataset and provide a baseline \CPSLint specification
    \item Filter unwanted substrings globally or per column
    \item Enforce data types on columns
    \item Enforce a valid range of values for numeric columns
    \item Impute missing data with configurable strategies
    \item Compartmentalise larger datasets based on flags
\end{itemize}

\CPSLint is implemented using the Rascal meta programming language~\cite{Klint:2009:RDSL, Klint:2011:EMPR}, a \enquote{one-stop-shop} language workbench that allows the rapid implementation of DSLs. Our implementation allows users to run \CPSLint code and contains two pipelines. One pipeline is a compiler that generates human-readable (and thus modifiable) Python code which performs the declared actions in the \CPSLint definition with the help of  Python  data science libraries. The second pipeline is an interpreter that performs the \CPSLint code within Rascal itself, providing high-granularity insights into the performed operations in the form of logs and intermediate results, at the cost of a performance penalty. 

The \texttt{inspect} action, depicted in \Cref{lst:inspect1}, instructs \CPSLint to analyse the input CSV file and generate a specification for the CSV in \CPSLint. Here, the column names and types are extracted with the help of \texttt{pandas}~\cite{McKinney:2010:DSSC} when the compiler is used, or with Rascal's built-in CSV parser when the interpreter is used. The fewer corruptions are in the data used for this inspection, the closer the inferred specification will be to the ideal. 
\begin{lstlisting}[
    caption={Example of the \texttt{inspect} action, used to generate a starting point for working with new data.}, 
    label={lst:inspect1}
]
inspect csv from 'input.csv';
\end{lstlisting}

The \texttt{import} action, depicted in \Cref{lst:import}, specifies which file will be used for the preprocessing and can immediately apply some global (file-wide) filters such as skipping empty lines and removing (sub)strings that appear through the data as a result of faulty readings or other corruptions.
\begin{lstlisting}[
    caption={Code for importing an input CSV, also specifying to skip empty rows as a preprocessing step.}, 
    label={lst:import}
]
import csv from 'data_input.csv', skip empty;
\end{lstlisting}

The \texttt{export} action allows the user to create a mapping between columns from the imported CSV and columns of the output CSV. For each column, the user can apply filters to enforce data types, set valid ranges for numerical values, and define strategies for imputing missing or corrupt data. The language also allows sequential application of the filters, making it more flexible in how the filters are applied. Besides exporting the sanitised data to a single file, it is also possible to cut the data while exporting based on a condition that is evaluated on the data in a column. An example of the export action is listed in \Cref{lst:export}.
\begin{lstlisting}[
    caption={Code for exporting an output CSV, mapping input to output CSV columns, specifying valid numeric ranges and imputation of empty cells for voltage using the mean.}, 
    label=lst:export
]
export csv
(
 'Timestamp (S)' is timestamp, in [0..50], critical;
 'Arc Main Cur (A)' is current, in [0.6..1.4];
 'Arc Main Vol (V)' is voltage, in [5±0.1], impute mean;
 'Arc Main Enr (J)' is energy, in [0..250];
 'Arc UART (TXT)' is 'UART';
)
to 'data_export_#.csv' cut when ' UART ' is ' image loader '~;
\end{lstlisting}

The current prototype supports table-wide sanitisation of empty rows and (sub)strings, data type enforcement on columns supporting elementary data types (such as \texttt{int}, \texttt{bool}, \texttt{real}) and domain-specific data types (such as \texttt{UART}), valid range enforcement on numeric columns, conditional sanitisation, imputation using different strategies (such as \texttt{mean}, copying the \texttt{last} or \texttt{next} valid values, various methods of \texttt{interpolation}), and phase detection through explicit flags a column. 

With these core features, \CPSLint can be used to prepare datasets for ML workflows. The current functionality of the language mostly focuses on the facilitation of remedies for corrupted data. For example, missing sensor data can be imputed using an appropriate technique, e.g., taking the mean of all available values, or performing interpolation. Remediation strategies are fairly configurable, e.g., the type of interpolation to be used.

\begin{figure*}[htbp]
	\centering
    \includegraphics[width=0.90\linewidth]{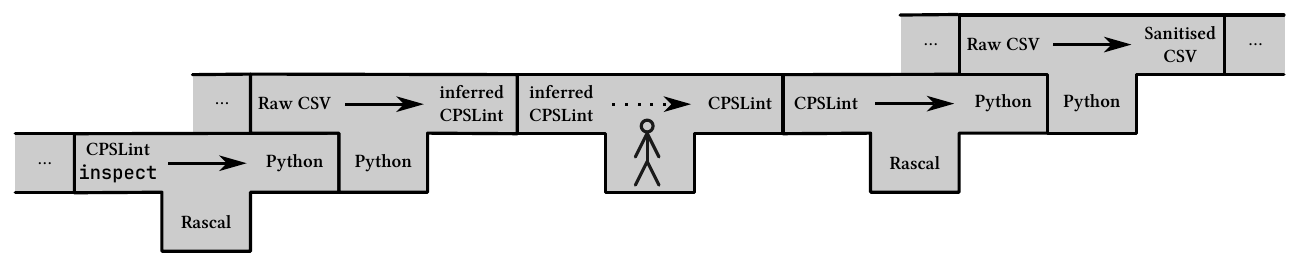}
	\caption{A tombstone diagram of the \CPSLint compiler pipeline. Activities flow rightwards, with inspection and inferring of the data structure happening on the left side, followed by Python code generation to deliver the executable code operating on the actual machine trace. Ellipses indicate where the normal data processing flow connects. Note the role of a domain expert refining the inferred \CPSLint specification.}
	\label{fig:cpslint_pipeline}
\end{figure*}

\paragraph*{The compiled pipeline}
This pipeline is used to transform \CPSLint code into executable Python code that can read an input dataset and clean it according to the strategies specified in the \CPSLint program, with the help of common libraries such as \texttt{pandas}, \texttt{numpy}, and \texttt{scipy}. the complete compilation to execution pipeline is depicted in \Cref{fig:cpslint_pipeline}.

On the left, our compiler written in Rascal receives the \texttt{inspect} command in \CPSLint and generates Python code that generates the first draft of a \CPSLint specification. The specification will be based on column names (explicit in the CSV header) and inferred data types. The domain expert, in the middle, further refines this specification. Moving to the right, the full \CPSLint specification is compiled to sanitisation code by our compiler. Finally, on the top right, the generated Python script takes in the same (or similar) raw CSV file as it was used at the start of this process, with the machine trace and possibly corrupt data. The Python script produces a sanitised CSV from this input according to the rules expressed in the \CPSLint specification. A truncated example of the Python code generated by the compiler pipeline of \CPSLint can be seen in \Cref{lst:pythonout}.
\begin{lstlisting}[
    caption={A truncated example of generated Python code that would be produced by running a program made from the contents of \Cref{lst:import} and \Cref{lst:export}.}, 
    label={lst:pythonout}, 
    language=PythonExtended
]
# Generated by CPSlint
# Input: specification.cps
import pandas as pd
# ...statically generated helper functions...
# Read input CSV
df_in = pd.read_csv(input_path)
# Apply global rules from specification
df_in = df_in.replace(r'^\s*$', pd.NA, regex=True).dropna(how="all")
# Create output DataFrame
df_out = pd.DataFrame()
# Copy/rename columns to the output DataFrame 
df_out["timestamp"] = df_in["Timestamp (S)"]
...
# Apply column specific rules 
df_out.loc[~df_out["timestamp"].between(0., 50., inclusive="both"), "timestamp"] = np.nan
...
# Write Output CSV
df_out.to_csv(output_path, index=False)
\end{lstlisting}

\begin{figure}[htbp]
	\centering
    \includegraphics[width=\linewidth]{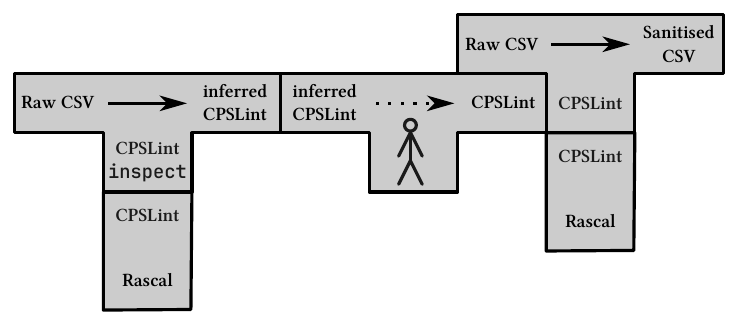}
	\caption{A tombstone diagram of the \CPSLint interpreter: the presence of the interpreter (vertical block) of \CPSLint written in Rascal allows us to see \CPSLint specifications as transformations from raw CSV to its sanitised form. }
	\label{fig:cpslint_pipeline_interpreter}
\end{figure}

\paragraph*{The interpreted pipeline}
The interpreter pipeline directly executes the \CPSLint specification using Rascal, relying on our runtime library and utilities, as well as functionality found in the Rascal standard library. This pipeline was built to enable the gathering of more granular insight into the data sanitisation process in the form of logging and saving intermediate results for inspection and debugging of the \CPSLint specification scripts. The details of this pipeline are depicted in \Cref{fig:cpslint_pipeline_interpreter}.

The workflow for the domain expert largely remains the same, with the main difference being that Python is now removed. Reading the diagram from left to right: the interpreter written in Rascal receives the \texttt{inspect} command and is instructed to read a raw CSV file and analyse its headers and the data types of the columns to generate the first draft of a \CPSLint specification. Then, in the same fashion as in the compiler pipeline, the domain expert refines this specification to describe the ideal form of the dataset. Finally, on the right, the interpreter receives the refined \CPSLint specification and is instructed to generate a sanatised CSV based on the strategies specified by the domain expert.

\subsection{Extensibility}
Future iterations of \CPSLint are planned to bring features that would better cater the language to the domain for which it is built. The declarative nature of the syntax of \CPSLint makes the language relatively easy to extend at the syntax level, with new features requiring only a few additional keywords in the grammar. The main additions to the language would introduce the concept of domain-specific units, based on SI~\cite{BIPM:2025:SI} and non-SI units, e.g., volt, watt, and mAh, as native data types. This would, for example, enable to perform simple type conversions on columns. Other extension broadening the applicability of the DSL, is support for more data source formats, e.g., HDF5 files or time-series databases. Future iterations of \CPSLint could support compartmentalisation based on numerical columns, where, for instance, time-series data is cut according to peaks, valleys, and plateaus in these values. Peak detction is advantagous where logged indicators cannot be relied upon.

Another extension that would make the language more broadly applicable is the support for custom Python libraries within the compiler pipeline. These would contain functions taking either a \texttt{DataFrame} (for operations at the table level) or a \texttt{Series} (for operations at the column level) as their input. In this way, users can add their own case-specific algorithms to preprocess data using \CPSLint. An example of what this extension could look like is depicted in \Cref{lst:pylibs}.
\begin{lstlisting}[
    caption={Syntax for using external libraries in \CPSLint.}, 
    label={lst:pylibs}
]
using <library.py>;
import csv from 'input.csv', skip empty, skip '#', skip '@', perform <tableOperation>;
export csv
( 
 ...
 'Arc Main Current (A)' is current, real, in [0.6..1.4], perform <rowOperation>;
 ... 
)
to 'output.csv';
\end{lstlisting}

\section{Demonstrator description}
\label{sec:demonstrator_description}
A complete use-case involving the concept of execution phases through a phase-based solution is chosen as the demonstration platform. The behaviour includes aforementioned traits expected from industrial CPS (\Cref{subsec:industrial_cyber_physical_systems}).

\subsection{Demonstration platform}
\label{subsec:demonstration_platform}
\begin{figure}[htbp]
	\centering
    \includegraphics[width=0.90\linewidth]{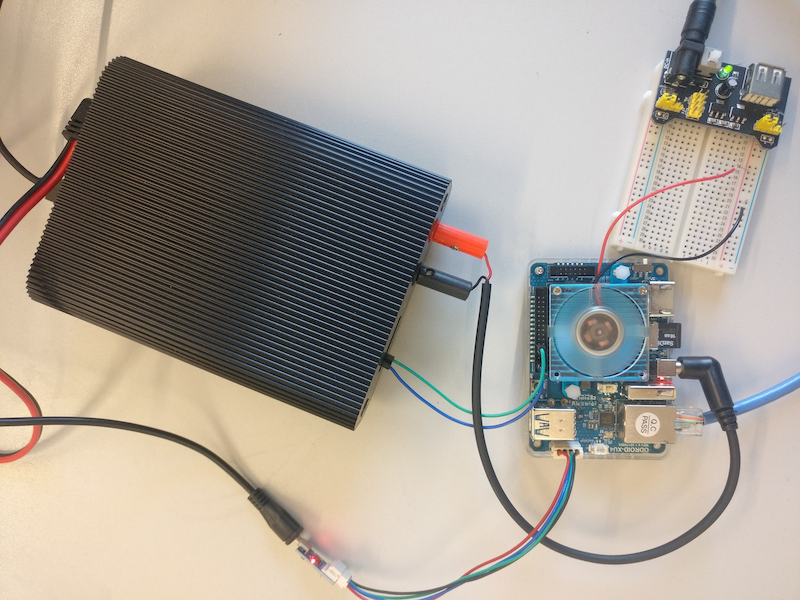}
	\caption{Experimental setup used as the demonstration platform, covering devices for electric power measurements.}
    \label{fig:experimental_setup}
\end{figure}

The reference monitoring data is collected from a computing device based on the ODROID-XU4, incorporating the ARM big.LITTLE architecture. The considered workload is an image processing application running on a Linux OS with minimal footprint. Images can be read either from the attached camera, or the available storage device. The monitoring device is an external power data logger, sitting between a high-precision Power Supply Unit (PSU) and the computing device. A separate PSU feeds the cooling fan to enforce isolation and to ensure correctness of readings~\cite{Odyurt:2021:PPFT}. Note that this is a real laboratory setup, as shown in \Cref{fig:experimental_setup}, producing real measured data; it is neither simulated nor synthetic.

\subsection{Reference data}
To collect the chosen reference monitoring data, the experimental platform has been extensively benchmarked~\cite{Odyurt:2021:PPFT}. Monitoring data is collected while running different workloads, i.e., the image processing application, running with different inputs, and a variety of operational conditions, resulting in 24 data collection scenarios. These are highly controlled experimental scenarios and are defined by combinations of the following parameters:
\begin{itemize}
    \item 3 physical execution conditions, \texttt{Normal}, \texttt{NoFan} (anomalous), \texttt{UnderVolt} (anomalous);
    \item 2 sets of workload inputs, i.e., different sets of images with 30 counts for each;
    \item 2 selected core types from the available hardware, Little Core and Big Core;
    \item 2 repetition counts, single round of input processing and 10 rounds of input processing.
\end{itemize}

This reference data covers electrical metrics (potential, current, power) in the form of time-series, augmented by core clock readings and multiple platform temperature sensor readings. Collection frequencies vary, in particular, electric potential and current are collected with the sampling rates of 1~kHz and 4~kHz, respectively. Lastly, messages sent over UART are included, acting as signalling data to follow individual tasks and sub-tasks.

\subsection{Raw data description}
Data collected directly from the logger contains the headers:
\begin{itemize}
    \item Timestamp (S)
    \item Arc Main Current (A)
    \item Arc Main Voltage (V)
    \item Arc Main Energy (J)
    \item Arc UART (TXT)
\end{itemize}

The column titles are self-explanatory. UART cells are not populated in every row. Useful UART messages are limited to start of processing, start and end of different activities (later interpreted as execution phases), core clock readings and board temperature readings, collected with the low frequency of per processed input instance.

\subsection{Emulating data corruptions}
To systematically evaluate the capabilities of \CPSLint, we have designed a set of corruption strategies, mimicking common errors observed in monitoring data, as described in \Cref{subsec:data_corruption_types}. These corruptions are applied to the original raw data in localised, non-overlapping blocks of rows (10 rows). The corruptions affect a predefined fraction of the total data, in this case 0.5\%. The motivation behind corrupting contiguous segments is to ensure that the effects are realistic, challenging and resembling real transmission errors, sensor failures, or logging issues. During high-frequency logging, trace logs usually portray a delay effect before going back to normal. Depending on the size of the original data, which is large in our case, there will be numerous 10-row blocks with a rate of 0.5\% injected corruptions. Note that the corruption rate has no effect on the ability to remedy, i.e., remediation works regardless of the amount. The following corruption types are supported and can be emulated:
\begin{itemize}
    \item Data type mismatch
    \item Data type mismatch with targeted UART
    \item Out-of-bounds values
    \item Out-of-order rows with reliable timestamps
    \item Out-of-order rows with unreliable timestamps
    \item Missing fields
    \item Missing rows
\end{itemize}

The algorithms emulating these strategies are implemented in Python. Any combination of the available corruption can be emulated at the same time. However, to keep experimentation and evaluate \CPSLint's capabilities, corruption emulations are executed individually on a reference trace file. A detailed \texttt{JSON} log file records the location of corrupted blocks within the reference trace file, alongside the emulated corruption type, which is highly useful for debugging and tracking of remedies later on.

\section{Application of \CPSLint}
\label{sec:application_of_cpslint}
The utilisation of \CPSLint can be divided into countering data corruptions and performing data compartmentalisation, i.e., associating traces with execution phases. Below, we cover some of the capabilities of \CPSLint using the dataset obtained from our demonstrator system (\Cref{sec:demonstrator_description}).

\subsection{Data inspection and script generation}
This feature of \CPSLint is used as a starting point for working with new data. The user provides \CPSLint alongside the selected data, asking to \enquote{inspect} it. This instructs \CPSLint to examine the provided dataset and perform type inference on the columns (illustrated in \Cref{lst:inspect2}). This type inference relies on the \texttt{dtypes} pandas assigns to a column based on the data in it in the compiled pipeline and on the datatypes inferred by \texttt{lang::csv} from the Rascal standard library in the interpreted pipeline.
\begin{lstlisting}[
    caption={Example of the \texttt{inspect} action, used to generate a starting point for working with new data.}, 
    label={lst:inspect2}
]
inspect csv from 'input.csv';
\end{lstlisting}
This generates a \CPSLint specification that can be used as a baseline for further sanitisation. For example, if the script from \Cref{lst:inspect2} is run on clean data, it generates the specification shown in \Cref{lst:generated}.
\begin{lstlisting}[
    caption={A script generated from analysed input data.}, 
    label={lst:generated}
]
import csv from 'input.csv', skip empty;
export csv
(
 'Timestamp (S)' is 'Timestamp (S)', real;
 'Arc Main Current (A)' is 'Arc Main Current (A)', real;
 'Arc Main Voltage (V)' is 'Arc Main Voltage (V)', real;
 'Arc Main Energy (J)' is 'Arc Main Energy (J)', real;
 'Arc UART (TXT)' is 'Arc UART (TXT)'; 
)
to 'output.csv';
\end{lstlisting}

\subsection{Countering data corruption}
An extensive list of potential corruptions has been introduced in \Cref{subsec:data_corruption_types}. Having the golden data, we systematically generated representative data for each type, as explained. Below we elaborate the use of \CPSLint in sanitising these corruption types.

\paragraph*{Data type mismatch (with targeted UART)}
One way to remedy this corruption is by applying global substring filters on the import action as shown in line~1 of \Cref{lst:remedymismatch}. Here, \texttt{skip regex \enquote*{[\#@\$]}} causes the substrings matching the regular expression \texttt{[\#@\$]} to be filtered out of the data, and similarly, \texttt{skip \enquote*{*}} causes occurrences of the character \texttt{\enquote*{*}} to be filtered. These substring filters are also applicable on specific columns, as illustrated in line 4. Finally, to make sure the columns have parsable entries for their columns, the desired data type can be enforced such that  entries that are not parsable as an instance of that data type are removed. For columns with numeric values, any rows that were emptied as an effect of the data type enforcement can be imputed, in the example this is done using linear interpolation.
\begin{lstlisting}[
    caption={A snippet to remedy mismatched data types.}, 
    label={lst:remedymismatch}
]
import csv from 'input.csv', skip regex '[#@$]', skip '*';
export csv
( 
 ...
 'Arc Main Current (A)' is current, skip '#', real, impute linear interpolation;
 ...
 'Arc UART (TXT)' is 'UART', uart;
)
to 'output.csv';
\end{lstlisting}

\paragraph*{Out-of-bounds values}
The remedy for this corruption is illustrated in \Cref{lst:oobv}. Here, we first enforce the data types of the columns to ensure that all columns for which we define a valid range contain only numeric values. Then, we can set a valid range for these column, which will cause cells with values outside of this defined range to be emptied. The notation here allows to specify what is included within the range, with rounded brackets meaning exclusion and square brackets meaning inclusion.  
\begin{lstlisting}[
    caption={A snippet demonstrating how to define a valid range for numeric data.}, 
    label={lst:oobv}
]
import csv from 'input.csv';
export csv
(
 'Timestamp (S)' is timestamp, real, in (0..50];
 'Arc Main Current (A)' is current, real, in [0.6..1.4);
 'Arc Main Voltage (V)' is voltage, real, in [5±0.1];
 'Arc Main Energy (J)' is energy, real, in (0..250);
 'Arc UART (TXT)' is 'UART', uart; 
)
to 'output.csv';
\end{lstlisting}

\paragraph*{Out-of-order rows with reliable timestamps}
For this corruption we can leverage the reliable timestamps and simply enforce the timestamps column to be sorted. This is illustrated in \Cref{lst:tsreliablesort}.
\begin{lstlisting}[
    caption={Sorting the data based on values in a column.}, 
    label={lst:tsreliablesort}
]
import csv from 'input.csv';
export csv
( 'Timestamp (S)' is timestamp, sorted; ... )
to 'output.csv';
\end{lstlisting}

Another way to tackle corruptions of this kind is to skip rows in the dataset that appear out of order. This method of sanitisation, illustrated in \Cref{lst:tsreliableskipooo}, is slightly more aggressive since it outright removes data and is meant to be used in scenarios where the previous method is not appropriate.  
\begin{lstlisting}[
    caption={A snippet skipping rows where the timestamp column is out of order.}, 
    label={lst:tsreliableskipooo}
]
import csv from 'input.csv';
export csv
(
 'Timestamp (S)' is timestamp, skip out of order; 
 ... 
)
to 'output.csv';
\end{lstlisting}

\paragraph*{Out-of-order rows with unreliable timestamps}
As the temporal dimension is the main reference point, which is not reliable, detecting and remedying is not trivial. The detection will rely on abnormal metric value progressions. At the same time, the analysis should ensure that the abnormal progression is unrelated to an actual anomaly resulting from machine behaviour. We leave it as future work.

\paragraph*{Missing fields}
The way this corruption can be tackled is highly dependent on the columns that have missing fields. For certain columns like voltage, imputing the missing values with the mean of the present values could be a good strategy, since the values have a valid range from 4.9 to 5.1. This is illustrated in \Cref{lst:mfv}.
\begin{lstlisting}[
    caption={\CPSLint Specification where rows where missing values get imputed using the mean of present values.}, 
    label={lst:mfv}
]
import csv from 'input.csv';
export csv
(
 ...
 'Arc Main Voltage (V)' is voltage, real, in [5±0.1], impute mean;
 ... 
)
to 'output.csv';
\end{lstlisting}
For columns like timestamp, where the values are expected to be increasing, imputing using linear interpolation is a good fit, since this takes the last and next known valid values into account, i.e., the imputed value(s) should not end up being larger that the proceeding correct row's timestamp. The snippet in \Cref{lst:mfts} contains the code to express this.
\begin{lstlisting}[
    caption={Example of how a missing timestamps can be imputed using linear interpolation.}, 
    label={lst:mfts}
]
import csv from 'log_input.csv';
export csv
( 
 'Timestamp (S)' is timestamp, real, impute linear interpolation; 
 ... 
)
to 'log_output.csv';
\end{lstlisting}

\paragraph*{Missing rows}
This corruption is particularly challenging to detect since we cannot say for certain how many rows of the data are missing, as the file itself will not contain any indication. One potential indication of such a corruption could be an unusual gap in the temporal progression of traces. Assuming detection, we can perform a similar remediation to missing fields, but performed at scale. For the numeric columns in the dataset, we can impute the missing values with the appropriate method per column. In the current iteration of \CPSLint we do not have a strategy for columns with other data types such as UART. An example of the code to remedy this corruption can be seen in \Cref{lst:missingrows}.
\begin{lstlisting}[
    caption={Example showing how multiple columns of missing data can be imputed at once in the case where missing rows are detected.}, 
    label={lst:missingrows}
]
import csv from 'input.csv';
export csv
(   
 'Timestamp (S)' is timestamp, real, in [0..50], impute linear interpolation;
 'Arc Main Current (A)' is current, real, in [0.6..1.4], impute last;
 'Arc Main Voltage (V)' is voltage, real, in [5±0.1], impute mean;
 'Arc Main Energy (J)' is energy, real, in (0..250), impute last;
 'Arc UART (TXT)' is UART, uart; 
)
to 'output.csv';
\end{lstlisting}

\subsection{Data compartmentalisation}
\CPSLint can also be used to cut the time-series data into shorter segments. The primary aim of cutting is to separate data associated with each defined execution phase. In the current implementation, data can be compartmentalised by observing substrings in the flags:
\begin{lstlisting}[
    caption={\CPSLint specification with which data is cut based on textual markers from a designated column.}, 
    label={lst:cutting}
]
import csv from 'input.csv', skip empty;
export csv
( ... )
to 'output#.csv'
   cut when 'UART' is 'image loader'~;
\end{lstlisting}

Here, the larger CSV file with time-series data is cut into numbered smaller files by cutting the data at points where an image loading phase happens. 

\section{Benchmarking setup}
\label{sec:benchmarking_setup}
One of the comparison dimensions considered for alternative remediation approaches, is computational performance. The metrics covered in this dimension are CPU-time\footnote{On a test system with no other running jobs, duration, i.e., wall-clock time, will be rather close to CPU-time.} and max Resident Set Size (Max-RSS), i.e., memory. The collection resolution is in nanoseconds and Bytes, respectively.

\begin{table}[t]
    \centering
    \caption{List of considered corruptions for benchmarking, alongside the applied remediation policies.}
    \label{tab:benchmarking_corruptions}
    \begin{tabularx}{\columnwidth}{@{}lX@{}}
        \toprule
        \textbf{Corruption type} & 
        \textbf{Policy} \\
        \midrule
        Data type mismatch & 
        Skip corrupted characters, enforce \texttt{real}-typed values in numeric columns, use linear interpolation to impute empty cells. \\
        Out-of-bounds values & 
        Set the data type to \texttt{real} in numeric columns, then enforce the data to lie within a valid range. \\
        Missing fields & 
        Set the data type to \texttt{real} in numeric columns, impute timestamps with linear interpolation, impute voltage with mean. \\
        Out-of-order rows\\with reliable timestamps & 
        Sort timestamps. \\
        Out-of-order rows\\with reliable timestamps & 
        Skip out-of-order timestamps. \\
        \bottomrule
    \end{tabularx}
\end{table}

\begin{table*}[t]
    \centering
    \caption{Computational performance results for: Pure Python (Py), Compiled \CPSLint (CPy), Interpreted \CPSLint (IC) on the five remedy types over ten runs. The measurements for Pure Python (Py) and Compiled \CPSLint (CPy) were collected using \texttt{rusage}, while the Interpreted \CPSLint (IC) was measured using Rascal's \texttt{util::Benchmark} library (lower is better).}
    \label{tab:benchmarking_results}
    \begin{tabular}{@{}lrrrrrrrrr@{}}
        \toprule
        \textbf{Remedy} &

        \multicolumn{3}{c}{\shortstack{\textbf{CPU-time mean (millisec)}}} &
        \multicolumn{3}{c}{\shortstack{\textbf{Max-RSS mean (MB)}}} \\
        
        \cmidrule(lr){2-4}
        \cmidrule(lr){5-7}

        & \multicolumn{1}{c}{\textbf{Py}} & \multicolumn{1}{c}{\textbf{CPy}} & \multicolumn{1}{c}{\textbf{IC}}
        & \multicolumn{1}{c}{\textbf{Py}} & \multicolumn{1}{c}{\textbf{CPy}} & \multicolumn{1}{c}{\textbf{IC}} \\
        \midrule
         Data type mismatch       & \textbf{1719.163} & 1937.295          & 253965.161 & 284.591 & \textbf{183.882} & N/A \\
         Out-of-bounds values     &  619.768          &  \textbf{401.043} & 292048.879 & 245.744 & \textbf{159.655} & N/A \\
         Missing fields           &  693.058          &  \textbf{541.538} & 129109.355 & 247.748 & \textbf{159.792} & N/A \\
         Out-of-order rows (skip) &  768.855          &  \textbf{444.752} &  77832.279 & 259.280 & \textbf{159.729} & N/A \\
         Out-of-order rows (sort) &  765.268          &  \textbf{453.281} &  56656.441 & 258.071 & \textbf{159.752} & N/A \\
        \bottomrule
    \end{tabular}
\end{table*}

\subsection{Benchmarking methodology}
There are four types of corruptions that are considered and applied to a single reference trace file, leading to four corrupted trace files. The applied corruptions and their corresponding remediation policies are listed in \Cref{tab:benchmarking_corruptions}. Note that the type \enquote{out-of-order rows with reliable timestamps} is dealt with twice: once in \enquote{skip out-of-order} mode and once in \enquote{sort} mode, as these can both be valid ways to remedy this type of corruption.

\begin{figure}[thbp]
	\centering
    \includegraphics[width=0.95\linewidth]{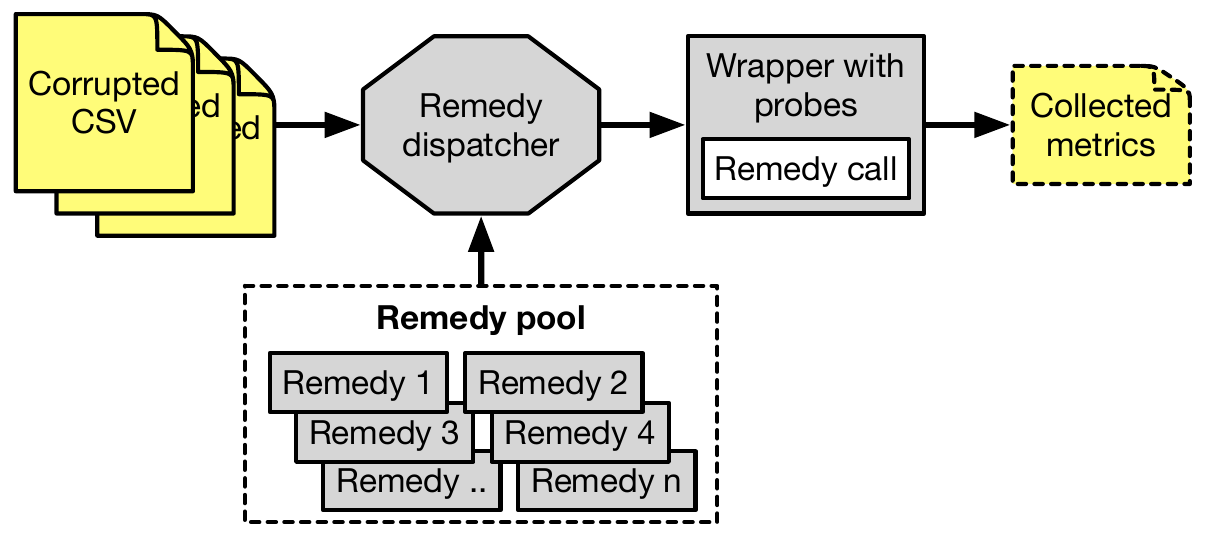}
	\caption{Benchmark collection setup, taking advantage of a Python wrapper around remediation calls to facilitate recording of computational performance metrics.}
	\label{fig:benchmarking_setup}
\end{figure}

To improve statistical rigour and to mitigate initialisation bias, i.e., cold-start effects, each remediation experiment is repeated 11 times and the first run is discarded as a warm-up. Reported results are based on the remaining 10 runs. The diagram from \Cref{fig:benchmarking_setup} visualises the benchmarking setup.

\subsection{The implementations}

The \textbf{\emph{Pure Python (Py)}} approach is representative of the current situation, in which Python code for data processing is written by a data scientist or a software engineer. There will be input from a domain expert. We have chosen to use only pure Python code with a minimal number of external libraries, as \emph{achieving minimal dependency surface is often the practice in the industry}. This approach will cover all possible implementations as well, i.e., pure Python, Pandas-based, and Rascal.

The \textbf{\emph{Compiled \CPSLint (CPy)}} is the scenario in which the compiled pipeline of \CPSLint is used, where the data specification written in \CPSLint code is transformed into Python code by our language implementation written in Rascal, which can then be executed to perform the preprocessing. For this experiment, we are not considering the step of processing the \CPSLint specification into Python code; rather, we consider the generated Python code itself.

The \textbf{\emph{Interpreted \CPSLint (IC)}} approach is largely similar to the compiled approach, in the sense that both of them require the user to provide a \CPSLint specification. Instead of generating Python code, this approach preprocesses the data directly using our Rascal implementation.

\subsection{Benchmark results}
The performance metrics obtained by benchmarking the three approaches are presented in \Cref{tab:benchmarking_results}. We report  mean CPU-time (in milliseconds) and memory usage (in megabytes) for each approach, while running remedies addressing the considered corruptions form \Cref{tab:benchmarking_corruptions}. Specific to Interpreted \CPSLint approach, we do not collect memory usage, as this approach relies on a completely different runtime (JVM) from the Python-based ones.


\section{Discussion}
\label{sec:discussion}
Considering the two approaches relying on Python, i.e., pure Python and compiled \CPSLint, we observe that in all but one case the compiled \CPSLint approach is faster than its pure Python counterpart. This speed-up can most likely be attributed to the data science libraries on which the compiled \CPSLint pipeline relies, i.e., Pandas, as these libraries are mature and well optimised. Another cause is algorithm design, e.g., row-wise processing vs column-wise processing and internal parallelism. Based on the memory usage comparison, we also observe that the Python code generated by \CPSLint is more efficient in this respect, which is again likely due to the efficient data structures and data reuse implemented by the underlying libraries.

Of all three options, the interpreted \CPSLint is evidently the slowest in all five cases. This was expected beforehand, since this approach implements the language in a way that focuses on gaining insight into the \CPSLint specification, logging every operation the and keeping intermediate CSVs after every operation. It is therefore heavily I/O bound. We believe that each approach has it's purpose:

The \textbf{\emph{Pure Python (Py)}} approach is the best pick if the main requirement is absolute control over the data sanitisation process, which may be preferred if performing processing unsupported by \CPSLint.

The \textbf{\emph{Compiled \CPSLint (CPy)}} approach provides a balance between control and convenience, i.e., users of \CPSLint can generate a Python script that processes the data according to their specification. These scripts can either be used directly or after modifications are made to the code.

The \textbf{\emph{Interpreted \CPSLint (IC)}} implementation is most advantageous for users who are already familiar with \CPSLint and aim to debug a specification under development. The performance penalty is otherwise too extreme to justify its use in production.

\section{Related work}
\label{sec:related_work}
At the time of writing, there exist several established tools and DSLs for data preprocessing. They usually support lightweight tabular transformations as well as expressive data querying and preprocessing.

In particular, \emph{GNU datamash}~\cite{GNU:2025:Datamash} is a popular command-line tool for basic numeric, textual, and statistical operations on tabular text data. Similar products complementing such shell-level workflows and leaning more towards embedded DSLs for batch processing and programmable querying, are: \emph{Lisp Query Notation}~\cite{Hoff:2024:LQND} which supports CSV, JSON and other kinds of structured data, in Common Lisp; \emph{Jet}~\cite{Ackermann:2012:JEDH} offers this for Scala and works with large datasets, generating optimised execution plans for Apache Spark and Hadoop.

Another line of work focuses on structuring preprocessing and pipeline specifications. \emph{Lavoisier}~\cite{DeLaVega:2020:LDIL} is a DSL for preparing datasets for data mining algorithms, reducing the complexity and size of preprocessing scripts compared SQL- or Pandas-based approaches; \emph{Papin}~\cite{Sal:2024:DSLA} extends it by integrating fishbone diagrams to explicitly model cause–effect relationships in data processing pipelines. For industrial deployment contexts, \emph{DSL4DPiFS}~\cite{Vogel-Heuser:2025:GNMD} offers a visual DSL for designing and deploying data pipelines in industrial metal forming systems, supporting optimisation, quality management, and predictive maintenance.

Some DSLs aim beyond transformation to make dataset properties explicit and checkable: \emph{DescribeML}~\cite{Giner-Miguelez:2023:DSLD} targets dataset documentation, describing structure, provenance, and social concerns, and works as a Langium-based VS Code extension; while \emph{RADAR}~\cite{Heine:2020:DADQ} targets data quality monitoring with integrity checks, null detection, and statistical validation through definitions of sources, checks, and actions.

In practice, much routine sanitisation is implemented manually on top of general purpose data analysis and learning libraries. \texttt{Pandas}~\cite{McKinney:2010:DSSC} provides foundational data structures and manipulation functionality for tabular data, making it probably the most common tool in Python-based pipelines. On top of it, \emph{scikit-learn}~\cite{Pedregosa:2011:Scikit} provides a broad suite of preprocessing operators and learning algorithms, and is often used for things like model-based screening of suspicious records. However, while these libraries make it easy to express transformations, they mostly treat data correctness as an application concern. Handling global formal guarantees about grammar/schema conformance, constraint satisfaction, as well as the main selling point of \CPSLint\ --- systematic handling of corruptions --- require not only explicit specifications, but also additional tooling.

A complementary line of research makes data assumptions explicit and mechanically checkable. Modern solutions require substantial investment but pay off by providing large scale services that incrementally grow with datasets~\cite{Schelter:2018:ALSD, Schelter:2019:UTDD}. In production ML pipelines, platform work such as \emph{TFX} (TensorFlow eXtended)~\cite{Baylor:2017:TFX} focuses on standardised components for analysing and validating both data and models. Beyond verification, the data management community has developed systems that not only detect, but also repair errors, often grounded in a taxonomy of common data quality issues and remedies~\cite{Rahm:2000:DCPC, Chu:2016:DCOE}. For example, \emph{NADEEF}~\cite{Dallachiesa:2013:NADEEF} provides an extensible system for data cleaning based on user-specified rules/constraints; \emph{Katara}~\cite{Chu:2015:KATARA} leverages knowledge bases (up to crowdsourcing) to interpret table semantics and suggest repairs; \emph{HoloClean}~\cite{Rekatsinas:2017:HoloClean} unifies constraint- and statistics-driven signals via probabilistic inference to produce holistic repairs at scale; and \emph{ActiveClean}~\cite{Krishnan:2016:ActiveClean} studies interactive/progressive cleaning tightly coupled to statistical model training, with convergence guarantees for the learning procedure.

While validation frameworks can be used in batch ingestion, operational settings often require continuous monitoring and alerting instead. Constraint-based verification can be extended into ongoing tracking of quality metrics and their changes over time~\cite{Schelter:2018:ALSD}. More recently, stream-first perspectives have been proposed that treat quality assessment itself as a continuous computation over unbounded data, producing quality \enquote{meta-streams} for pipeline awareness~\cite{Papastergios:2025:StreamDaQ}. In the context of industrial settings, proposals often explicitly integrate profiling, validation, and continuous monitoring as first-class pipeline stages under data quality dimensions~\cite{Peixoto:2025:DQPI}.


As a final remark, our architecture, with both an interpreter and a compiler for the same language, is not uncommon. This approach dates back to 4GLs such as Application Factory (later CorVision)~\cite{Martin:1986:FGL}, where users compile prototypes to high-level BUILDER code, debug via inspection or runtime interpretation, and, once stable, compile to deployable machine code. We envision a similar workflow: experimentation within Rascal is more efficient, while processing large data volumes and integrating data sanitisation into the data flow is better achieved with Python scripts.

\section{Conclusion and future work}
\label{sec:conclusion}

We introduced \CPSLint as a DSL for preparing machine data for data-centric workflows, such as anomaly detection and identification. We demonstrated its capabilities through examples, particularly its ability to compartmentalise data according to system execution phases. \CPSLint has two backends~\cite{Odyurt:2026:ImplementingCPSLint}: one generates Python code for orchestration alongside or prior to data-centric solutions, and the other directly interprets the specification using its Rascal implementation for debugging, with high-granularity logs and intermediate outputs. We evaluated both pipelines against each other and a pure Python implementation, comparing resource usage in terms of CPU time and memory.

As introduced in \Cref{sec:introduction}, a natural question arises: Why use a DSL~\cite{Mernik:2005:WHDD, Tomassetti:2020:RLAD} rather than implementing the functionality directly within the targeted solution? In this context, a DSL offers several advantages.

\textbf{Separation of concerns:}
As noted in \Cref{subsec:data_preprocessing}, preprocessing involves compartmentalising machine data according to the use-case. Each use-case has a distinct set of execution phases, whereas the targeted solutions consuming phase data, such as FDI workflows, remain unchanged. It is therefore advantageous to separate these concerns and delegate dynamic aspects of machine data to a descriptive tool which can be individually reconfigured. Beyond phase definitions, expected corruptions could also be considered dynamic, for instance, depending on machine type or model.

\textbf{Declarative expressiveness:}
The expressiveness of a DSL, being closer to domain language, enables this reconfigurability. It facilitates the involvement of domain experts who may not be programmers.

\textbf{Standardisation and consistency:}
A DSL enforces a contained and consistent set of rules, providing a stronger alternative to diverse programming styles in a general-purpose language. This is particularly beneficial in larger organisations where tasks are distributed among different engineers. In other words, the use of general-purpose languages by different programmers often results in ad hoc solutions. In contrast, a specialised DSL with a limited vocabulary ensures that the generated Python (or other) code employs embedded methodical techniques for supported functionality, such as filtering unwanted substrings. This methodical approach applies to both sanitisation and remediation techniques.

\textbf{On demand code generation:}
An additional advantage of \CPSLint, specifically in Python code generation mode, is the specificity of the generated code in relation with the task at hand, i.e., its \emph{on demand} nature. Although \CPSLint can be extended to support extensive catalogues of sanitisation and remediation techniques, individual use-cases require only a subset. The generated Python code therefore includes only the necessary functionality, avoiding the overhead of a maximalist package designed to serve all use-cases and resulting in a more streamlined approach.

\textbf{Future work --- Industrial case study:}
In \Cref{fig:machine_cycle}, we presented the high-level sequence of tasks, namely the machine cycle, for an ITEC die bonder machine. As industrial CPS sensor data is consistent in nature, our immediate future work is to apply \CPSLint to real ITEC machine data. This may require enhancements to address machine-specific corruptions and to compartmentalise data according to ITEC-specific execution phases, in collaboration with domain experts.

\begin{acks}
This publication is part of the project ZORRO\footnote{\url{https://zorro-project.nl}} with project number KICH1.ST02.21.003 of the research programme Key Enabling Technologies (KIC), which is (partly) financed by the Dutch Research Council (NWO).
\end{acks}


\balance

\bibliographystyle{ACM-Reference-Format}
\bibliography{bibliography/references}

\end{document}